\documentclass[journal,11pt]{IEEEtran}
\usepackage{algorithmic,algorithm,epsfig, amsmath,amssymb,amsfonts,cite,graphicx}
\usepackage{multirow}  
\newcommand{\comment}[1]{}

\newtheorem{theorem}{\bf Theorem}

\newtheorem{corollary}{\bf Corollary}

\newcommand{\qed}{\rule{2mm}{2mm}\bigskip}

\newcommand{\xin} {\mathsf{x}_{in}} 
\newcommand{\xout} {\mathsf{x}_{out}} 

\newcommand{\size}{\mathcal{M}}

\begin{document}
\title{A Survey on Network Codes \\ for Distributed Storage}
\author{Alexandros G. Dimakis,~\IEEEmembership{Member,~IEEE},
Kannan Ramchandran,~\IEEEmembership{Fellow,~IEEE},
Yunnan~Wu,~\IEEEmembership{Member,~IEEE}, Changho Suh,~\IEEEmembership{Student Member,~IEEE}
\thanks{Alexandros G. Dimakis with the Department of Electrical Engineering, University of Southern California, Los Angeles, CA 90089-2560.
{\tt dimakis@usc.edu.}}%
\thanks{Kannan Ramchandran and Changho Suh are with the Department of Electrical Engineering and Computer Science,
University of California, Berkeley, CA 94704. {\tt \{kannanr,chsuh\}@eecs.berkeley.edu}}%
\thanks{Yunnan Wu is with Microsoft Research, One Microsoft Way, Redmond, WA,
98052. {\tt yunnanwu@microsoft.com}.}}

\maketitle
\begin{abstract}
Distributed storage systems often introduce redundancy to increase reliability.
When coding is used, the \emph{repair problem} arises: if a node storing encoded information fails, in order to maintain the same level of reliability we need to create encoded information at a new node. This amounts to a partial recovery of the code, whereas conventional erasure coding focuses on the complete recovery of the information from a subset of encoded packets. The consideration of the repair network traffic gives rise to new design challenges. Recently, network coding techniques have been instrumental in addressing these challenges, establishing that maintenance bandwidth can be reduced by orders of magnitude compared to standard erasure codes. This paper provides an overview of the research results on this topic.
\end{abstract}

\section{Introduction}
In recent years, the demand for large scale data storage has increased significantly, with applications like social networks, file, and video sharing demanding seamless storage, access and security for massive amounts of data. When the deployed storage nodes are individually unreliable, as is the case in modern
data centers and peer-to-peer networks, redundancy must be introduced into the system to improve reliability against node failures. The simplest and most
commonly used form of redundancy is straightforward replication of the data in multiple storage nodes. However, erasure coding techniques can potentially achieve
orders of magnitude more reliability for the same redundancy compared to replication (see e.g.\cite{WeatherspoonK:02}). To realize the increased reliability of coding however, one has to address the challenge of maintaining an erasure encoded representation.

\begin{figure}[t!]
\centering
\vspace{0.5cm}
\includegraphics[width=8.5cm]{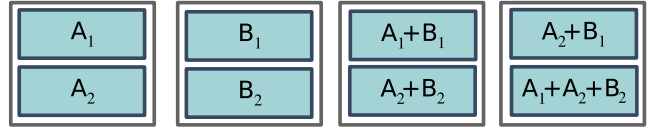}
\vspace{-0.5cm}
\caption{A (4,2) MDS binary erasure code (Evenodd Code~\cite{evenodd}). Each storage node (box) is storing two blocks that are linear binary combinations of the original data blocks $A_1,A_2,B_1,B_2$.
In this example the total stored size is $\size=4$ blocks.
Observe that any $k=2$ out of the $n=4$ storage nodes, contain enough information to recover all the data.}
\label{fig_ex1}
\end{figure}

Given two positive integers $k$ and $n>k$, an $(n,k)$ maximum distance separable (MDS) code can be used for reliability: initially the data to be stored is separated into $k$ information packets. Subsequently, using the MDS code, these are encoded into $n$ packets (of the same size) such that \emph{any} $k$ out of these $n$ suffice to recover the original data (see Figure~\ref{fig_ex1} for an example).

MDS codes are optimal in terms of the redundancy-reliability tradeoff because $k$ packets contain the minimum amount of information required to recover the original data. In a distributed storage system the $n$ encoded packets are stored at different storage nodes (e.g., disks, servers or peers) spread over a network, and the system can tolerate any $(n-k)$ node failures without data loss. Note that throughout this paper we will assume a storage system of $n$ storage nodes that can tolerate $(n-k)$ node failures and use the idea of sub-packetization: each storage node can store multiple sub-packets that will be referred to as blocks (essentially using the idea of array codes~\cite{evenodd,array1}).

The benefits of coding for storage are well known and there has been a substantial amount of work in the area. Reed--Solomon codes~\cite{ReedSolomon} are perhaps the most popular MDS codes and together with the very similar information dispersal algorithm (IDA)~\cite{Rabin}, have been investigated in distributed storage applications (e.g.\cite{Kubiatowicz+:00,bhagwan04total}). Fountain codes~\cite{Raptor} and LDPC codes~\cite{mct} are recent code designs that offer approximate MDS properties and fast encoding-and-decoding complexity. Finally there has been a large body of related work on codes for RAID systems and magnetic recording (e.g. see~\cite{evenodd,array1,array2,SiegelBook} and references therein).

\begin{figure}[t]
\vspace{0.3cm}
\includegraphics[width=8.5cm]{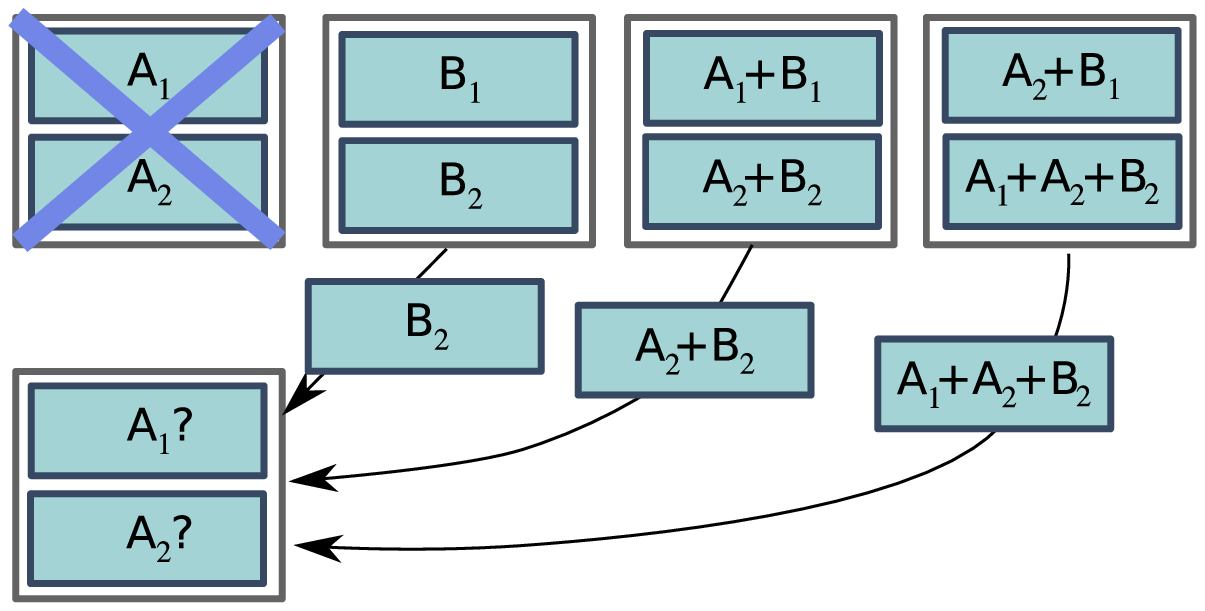}
\caption{Example of an (exact) repair: Assume that the first node in the previous storage system failed.
The question is to repair the failure by creating a new node (the newcomer) that still forms a (4,2) MDS code. In this example it is possible to obtain exact repair by communicating $3$ blocks, which is the information theoretic minimum cut-set bound.}
\label{fig_ex2}
\vspace{0.3cm}
\includegraphics[width=8.5cm]{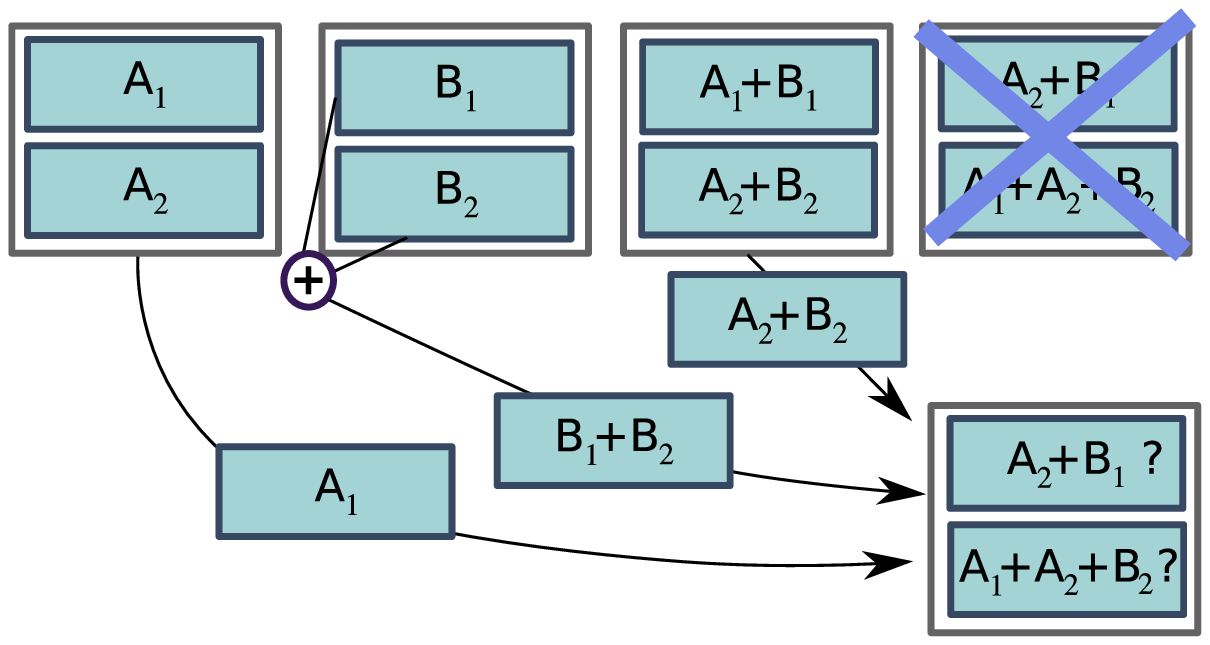}
\caption{Repairing the last node: in some cases it is necessary for storage nodes to compute functions
of their stored data before communicating, as shown in the second node.}
\vspace{-0.5cm}
\label{fig_ex3}
\end{figure}

In this tutorial we focus on a new problem that arises when \emph{storage nodes are distributed and connected in a network}. The issue of \emph{repairing a code} arises when a storage node
of the system fails. The problem is best illustrated through the example of Figure~\ref{fig_ex2}:
Assume a file of total size $\size=4$ blocks is stored using the $(4,2)$ Evenodd code of the previous example and the first node fails. A new node (to be called the newcomer) needs to construct and store two new blocks so that the three existing nodes combined with the newcomer still form a $(4,2)$ MDS code. We call this the \emph{repair problem} and focus on the required repair bandwidth. Clearly, repairing a single failure is easier than reconstructing all the data: since by assumption any two nodes contain enough information to recover all the data, the newcomer could download $4$ blocks (from any two surviving nodes), reconstruct all four blocks and store $A_1,A_2$. However, as the example shows, it is possible
to repair the failure by communicating only three blocks $B_2,A_2+B_2,A_1+A_2+B_2$ which can be used to solve for $A_1,A_2$.

Figure~\ref{fig_ex3} shows the repair of the fourth storage node. This can be achieved by using only three blocks~\cite{ITA_arrayrepair} but one key difference is that the second node needs to compute a linear combination of the stored packets $B_1,B_2$ and the actual communicated block is $B_1+B_2$.
This shows clearly the necessity of \emph{network coding}, creating linear combinations in intermediate nodes during the repair process.
 If the network bandwidth is more critical resource compared to disk access, as is often the case, an important consideration is to find what is the minimum required bandwidth and which codes can achieve it.

The repair problem and the corresponding regenerating codes were introduced in~\cite{DimakisGWWR:08} and received some attention in the recent literature~\cite{WuDR:07,Wu:09,Wu:10,WuD:09,CullinaDH:09,KVSKR:09,ShahRKR:09,SuhR:09,Wu:09c,
Dum09,LiInfocom10}. Somehow surprisingly these new code constructions can achieve a rather significant reduction in repair network bandwidth, compared with the straightforward application of Reed--Solomon or other existing codes. In this paper we provide an overview of this recent work and discuss
several related research problems that remain open.

\subsection{Various Repair Models}

In the repair examples shown in Figures~\ref{fig_ex2} and~\ref{fig_ex3}, the newcomer constructs exactly the two blocks that were in failed nodes. Note however that our definition of repair only requires that the new node forms an $(n,k)$ MDS code property (that any $k$ nodes out of $n$ suffice to recover the original whole data), when combined with existing nodes. In other words, the new node could be forming new linear combinations that were different from the ones in the lost node; a requirement that is strictly easier to satisfy.

\begin{figure}[t]
  \centering
  \includegraphics[width=8cm]{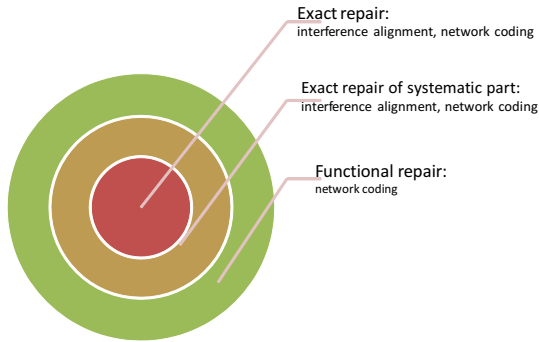}\\
  \caption{Various repair models and the key constructive techniques.}\label{fig:models}
\end{figure}

Three versions of repair have been considered in the literature: \emph{exact repair}, \emph{functional repair}, and \emph{exact repair of systematic parts}. In exact repair, the failed blocks are exactly regenerated, thus restoring exactly the lost encoded blocks with their exact replicas.
In functional repair, the requirement is relaxed: the newly generated blocks can contain different data from that of the failed node as long as the repaired system maintains the MDS-code property. The exact repair of the systematic part is a hybrid repair model lying between exact repair and functional repair. In this hybrid model, the storage code is always a systematic code (meaning that one copy of the data exists in uncoded form). The systematic part is exactly repaired upon failures and the non-systematic part follows a functional repair model where the repaired version may be different from the original copy. See Figure~\ref{fig:models} for an illustration. Notice that we do not know if the repair bandwidth
for the three cases can be made equal or not (so the subsets are not necessarily strict).

There is one important benefit in keeping the code in systematic form: as shown in Figure~\ref{fig_ex1},
if the code contains the original data as a subset, reading parts of the data can be performed
very quickly by just accessing the corresponding storage node without requiring decoding.
Interestingly, as we will see, exact repair which is the most interesting problem in practice, is also
the most challenging one and determining a large part of the achievable region remains open.

The functional repair problem is completely understood because as shown in~\cite{DimakisGWWR:08},
it can be reduced to a multicasting problem on an appropriately constructed graph
called the information flow graph.
The pioneering work of Ahlswede et al.~\cite{Ahlswede00} characterized the multicasting rates
by showing that cut-set bounds are achievable. Further work showed that linear network coding suffices~\cite{LYC03,KM03} and random linear combinations construct good network codes with high probability~\cite{RLNC}. See also the survey~\cite{Fragouli_primer} and references therein. Since functional repair is reduced to multicasting, we can completely characterize the minimum repair bandwidth by evaluating the min-cut bounds and network coding provides effective and constructive solutions. In Section~\ref{sec:functional} we present the results that characterize the achievable functional repair region and show a tradeoff between storage and repair bandwidth.

The exact repair problem is harder than the functional repair problem. In exact repair, the new node accesses some existing storage nodes and exactly reproduces the lost coded blocks. As will be described in the sequel, repair codes come with fundamental tradeoffs between storage cost and repair bandwidth. The two important special cases involve operating points corresponding to  maximal storage and  minimal bandwidth versus  minimal storage with maximal bandwidth point.  Exact repair for the minimal bandwidth operating point is described in  Section~\ref{sec:TwoSpecialCases}) and describes the recent work of ~\cite{KVSKR:09} which develops optimal exact repair codes for this operating point without any loss of optimality with respect to only functional repair.

The special case of the operating point that corresponds to minimal storage, which also corresponds to minimizing the repair bandwidth while keeping the same storage cost of MDS codes turns out to be more challenging. It turns out that in this case, the new node needs to recover part of the data which is \emph{interfered} with by the other data.
It is the need to carefully handle interference that makes the problem difficult.
The constructive techniques perform algebraic alignment so that the effective dimension of unwanted information is reduced, thus reducing the repair traffic. These constructive techniques building on the known alignment concept characterize the repair bandwidth for low-rate codes ($k/n \leq 1/2$) and constitute
achievable schemes for all the range of parameters. It remains however open if the cut-set bounds
are achievable for the whole range of parameters.

The exact repair of systematic parts model is a relaxation of the exact repair model. As in the exact repair model, the core constructive techniques are interference alignment and network coding. In Section~\ref{sec:hybrid}, we shall see that this relaxation addresses some problem space not covered by exact repair.

\section{Model I: Functional Repair}

\label{sec:functional}
\begin{figure}
    \centering
  \includegraphics[width=8cm]{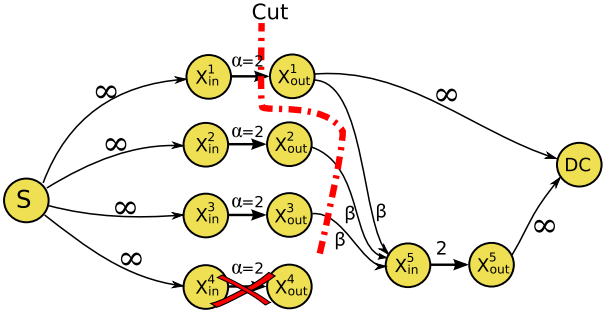}\\
  \caption{
Illustration of the information flow graph $\mathcal{G}$ corresponding to the (4,2) code of Figure 1. A distributed storage scheme uses an $(4,2)$ erasure
code in which any $2$ nodes suffice to recover the original
data. If node $x^4$ becomes unavailable and a new node joins the
system, we need to construct new encoded blocks in $x^5$. To do
so, node $x_{in}^5$ is connected to the $d=3$ active storage nodes.
Assuming $\beta$ bits communicated from each active storage
node, of interest is the minimum $\beta$ required.  The min-cut
separating the source and the data collector must be larger than $\size=4$ blocks
for regeneration to be possible.  For this graph, the min-cut value
is given by $\alpha+2\beta$, implying that communicating $\beta \geq 1$ block is sufficient and necessary. The total repair bandwidth to repair one failure is therefore $\gamma=d \beta= 3$ blocks.}
\label{fig:informationflowgraph}
\end{figure}

As shown in \cite{DimakisGWWR:08}, the functional repair problem can be represented as multicasting over an \emph{information flow graph}. The {\em information flow graph} represents the evolution of information flow as nodes join and leave the storage network (see also~\cite{Jiang06} for a similar construction). Figure~\ref{fig:informationflowgraph} gives an example information flow graph. In this graph, each storage node is represented by a pair of nodes, $\xin^i$ and $\xout^i$, connected by an edge whose capacity is the storage capacity of the node. There is a virtual source node $s$ corresponding to the origin of the data object. Suppose initially we store a file of size $\size=4$ blocks at four nodes, where each node stores $\alpha=2$ blocks and the file can be reconstructed from any $2$ nodes. Virtual sink nodes called \emph{data collectors} connect to any $k$ node subsets and ensure that the code has the MDS property (that any $k$ out of $n$ suffices to recover). Suppose storage node $4$ fails, the goal is to create a new storage node, node $5$, which communicates the minimum amount of information and then stores $\alpha=2$ blocks. This is represented in Figure~\ref{fig:informationflowgraph} by the unit-capacity edges $\xout^1 \xin^5$, $\xout^2 \xin^5$, and $\xout^3 \xin^5$ that enter node $\xin^5$.

The functional repair problem for distributed storage can be interpreted as a multicast communication problem defined over the information flow graph, where the source $s$ wants to multicast the file to the set of all possible data collectors. For multicasting, it is known that the maximum multicast rate is equal to the minimum-cut capacity separating the source from a receiver and it can be achieved using linear network coding \cite{LYC03}. Since the current problem can be viewed as a multicast problem, the fundamental limit can be characterized by the min-cuts in the information flow graph and network coding provides effective constructive solutions. One complication is that since the number of failures/repairs is unbounded, the resulting information flow graph can grow unbounded in size. Hence we have to deal with cuts, flows, and network codes in graphs that are potentially infinite.

In Section~\ref{sec:cut}  we present the cut analysis of information flow graphs \cite{DimakisGWWR:08,WuDR:07}. In Section~\ref{sec:TwoSpecialCases}, we discuss two extreme points corresponding to minimum repair bandwidth and minimum storage cost, respectively (arguably interesting cases).


\subsection{Cut Analysis of Information Flow Graphs}
\label{sec:cut}

By analyzing the connectivity in the information flow graph, we can derive fundamental performance bounds about codes. In particular, if the minimum cut between $s$ and a data collector is less than the size of original file, then we can conclude that it is impossible for the data collector to reconstruct the original file.
In this section we review the cut analysis of \cite{DimakisGWWR:08,WuDR:07}.
The setup is as follows: there are always $n$ active storage nodes. Each node can store $\alpha$ bits.
An information flow graph (as illustrated by Figure~\ref{fig:informationflowgraph}) corresponds to a particular evolution of the storage system after a certain number of failures/repairs. We call each failure/repair a ``stage"; in each stage, a single storage node fails and the code gets repaired by downloading $\beta$ bits each from any $d$ surviving nodes. Therefore the total repair bandwidth is $\gamma= d \beta$.

See Figure~\ref{fig:informationflowgraph} for an example. In the initial stage, the system consists of nodes $1,2,3,4$; in the second stage, the system consists of nodes $2,3,4,5$. For each set of parameters $(n,d,\alpha,\gamma= d \beta)$, there is a family of finite or infinite information flow graphs, each of which corresponds to a particular evolution of node failures/repairs. We denote this family of directed acyclic graphs by $\mathcal{G}(n,d,\alpha, \gamma)$.
We restrict our attention to the symmetric setup where it is required that any $k$ storage nodes can recover the original file, and a newcomer receives the same amount of information from each of the existing nodes.
 An $(n,k,d,\alpha,\gamma)$ tuple will be feasible, if a code with storage $\alpha$ and repair bandwidth $\gamma$ exists.
For the example in Figure~\ref{fig_ex2}, the total file has size $\size=4$ blocks and the point $(n=4,k=2,d=3,\alpha=2\,\text{blocks},\gamma=3\,\text{blocks})$ is feasible. On the contrary, a standard erasure code which communicates the whole data object would correspond to $\gamma=4\, \text{blocks}$ instead. Note that $n,k,d$ must be integers. If there is one failure, the newcomer can connect to at most to all the $n-1$ surviving nodes, so $d\leq n-1$ and $\alpha,\beta,\gamma=d \beta$ are the non-negative real valued parameters of the repair process.

\begin{theorem}
	\label{func_Repair_theorem}
For any $\alpha \geq \alpha^*(n,k,d,\gamma)$, the points $(n,k,d,\alpha,\gamma)$ are feasible and linear network codes suffice to achieve them. It is information theoretically impossible to achieve points with $\alpha<\alpha^*(n,k,d,\gamma)$.
The threshold function $\alpha^*(n,k,d,\gamma)$ is the following:
\begin{align}
\alpha^*(n,k,d,\gamma)=\left\{
                     \begin{array}{ll}
                       \frac{\size}{k}, & \gamma\in [f(0),+\infty) \\
                       \frac{\size-g(i)\gamma}{k-i}, & \gamma\in [f(i),f(i-1)),
                     \end{array}
                   \right.
\end{align}
where
\begin{align}
f(i)&\stackrel{\Delta}{=} \frac{2 \size d}{(2k-i-1)i+2k(d-k+1)},\label{eq:f_i}\\
g(i)&\stackrel{\Delta}{=} \frac{(2d-2k+i+1)i}{2d}\label{eq:g_i},
\end{align}
where $d \leq n-1$. Given $(n,k,d)$, the minimum repair bandwidth $\gamma$ is
\begin{align}
\gamma_{\min} = f(k-1) = \frac{2 \size d}{2kd-k^2+k}.
\end{align}
\end{theorem}
\vspace{0.5cm}

\begin{figure*}
\centering
\includegraphics[width=13cm]{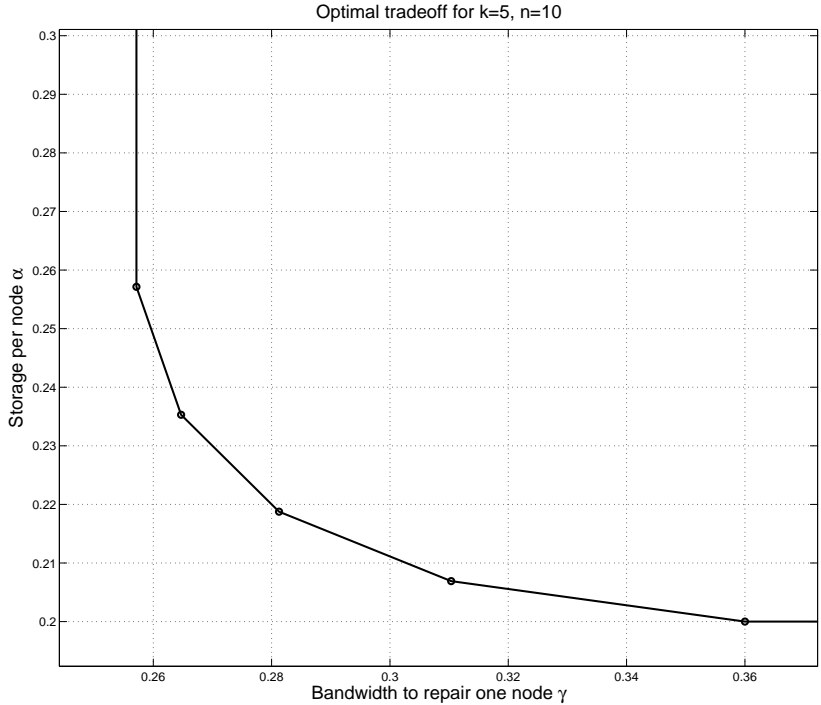}\\
\caption{Optimal tradeoff curve between storage $\alpha$ and repair bandwidth $\gamma$, for $k=5, n=10$. Here $\size=1$ and $d=n-1$.
Note that traditional erasure coding corresponds to the point $(\gamma=1,\alpha=0.2)$.
}
\label{fig:optimaltradeoff}
\end{figure*}

One important observation is that the minimum repair bandwidth $\gamma=d \beta$ is a decreasing function of the number $d$ of nodes that participate in the repair. While the newcomer communicates with more nodes, the size
of each communicated packet $\beta$ becomes smaller fast enough to make the product $d\beta$ decrease. Therefore, the minimum repair bandwidth can be achieved when $d=n-1$.

As we mentioned, code repair can be achieved if and only if the underlying information flow graph has sufficiently large min-cuts. This condition leads to the
repair rates computed in Theorem ~\ref{func_Repair_theorem}, and when these conditions are met,   simple random linear combinations will suffice with high probability as the field size over which coding is performed grows, as shown by Ho. et al.~\cite{RLNC}.
The optimal tradeoff curve for $k=5, n=10, d=9$ 
is shown in Figure~\ref{fig:optimaltradeoff}.

%

\subsection{Two Special Cases}
\label{sec:TwoSpecialCases}
It is of interest to study the two extremal points on the optimal tradeoff curve, which correspond to the best storage efficiency and the minimum repair bandwidth, respectively. We call codes that attain these points minimum-storage regenerating (MSR) codes and minimum-bandwidth regenerating (MBR) codes, respectively.

From Theorem~\ref{func_Repair_theorem}, it can be verified that the minimum storage point is achieved:
\begin{align}
\label{eq:MSRpoint}
(\alpha_{MSR}, \gamma_{MSR})=\left(\frac{\size}{k},\;\; \frac{\size d}{k(d-k+1)}\right).
\end{align}
As discussed, the repair bandwidth $\gamma_{MSR}=d \beta_{MSR}$ is a decreasing function of the number of nodes $d$ that
participate in the repair.
Since the MSR codes store $\frac{\size}{k}$ bits at each node while ensuring the MDS-code property, they are equivalent to standard MDS codes.
Observe that when $d=k$, the total communication for repair is $\size$ (the size of the original file). Therefore, if a newcomer is allowed to contact only $k$ nodes, it is inevitable to download the whole data object to repair one new failure and this is the naive repair method that can be performed for any MDS codes.

However, allowing a newcomer to contact more than $k$ nodes, MSR codes can reduce the repair bandwidth $\gamma_{MSR}$, which is minimized when $d=n-1$:
\begin{align}
	\label{MSR_cutset}
(\alpha_{MSR}, \gamma^{min}_{MSR})=\left(\frac{\size}{k},\;\; \frac{\size}{k} \cdot \frac{n-1}{n-k}\right).
\end{align}
We have separated the $\size/k$ factor in $\gamma^{min}_{MSR}$ to illustrate that MSR codes communicate an $\frac{n-1}{n-k}$ factor more than what they store. This represents a fundamental expansion necessary for MDS constructions that are optimal on the reliability-redundancy tradeoff. For example, consider a $(n,k) = (14,7)$ code. In this case, the newcomer needs to download only $\frac{\size}{49}$ bits from each of the $d= n-1 = 13$ active storage nodes, making the repair bandwidth equal to $\frac{\size}{7} \cdot \frac{13}{7}$. Notice that we need only an expansion factor of $\frac{13}{7}$, while a factor of 7 is required for the native repair method.


At the other end of the tradeoff are MBR codes, which have minimum repair bandwidth. It can be verified that the minimum repair bandwidth point is achieved by
\begin{align}
		\label{MBR_cutset}
(\alpha_{MBR}, \gamma_{MBR})=\left(\frac{2\size d}{2kd-k^2+k},\;\; \frac{2\size d}{2kd-k^2+k}\right).
\end{align}
Note that the minimum bandwidth regenerating codes, the storage size $\alpha$ is equal to $\gamma$, the total number of bits communicated during repair. If we set the optimal value $d=n-1$, we obtain
\begin{align}
\label{MBR_cutset_d_n-1}
(\alpha^{min}_{MBR}, \gamma^{min}_{MBR})=\left(\frac{\size}{k} \cdot \frac{2n-2}{2n-k-1},\;\;
\frac{\size}{k} \cdot \frac{2n-2}{2n-k-1} \right).
\end{align}
Notice that $\alpha^{min}_{MBR} = \gamma^{min}_{MBR}$: MBR codes incur no repair bandwidth expansion at all, just like a replication system does, downloading exactly the amount of information stored during a repair. However, MBR codes require an expansion factor of $\frac{2n-2}{2n-k-1}$ in the amount of stored information and are no longer optimal in terms of their reliability for the given redundancy.

\section{Model II: Exact Repair}
\label{sec:exactrepair}

As we discussed, the repair-storage tradeoff for functional repair can be completely characterized by analyzing the cut-set of the information flow graphs.
However, as mentioned earlier, functional repair is of limited practical interest since there is a need to maintain the code in systematic form. Also, under functional repair, significant system overhead is incurred in order to continually update repairing-and-decoding rules whenever a failure occurs. Moreover, the random network coding based solution for the function repair can require a huge finite-field size to support a dynamically expanding graph size (due to continual repair). This can significantly increase the computational complexity of encoding-and-decoding. Furthermore, functional repair is undesirable in storage security applications in the face of eavesdroppers. In this case, information leakage occurs continually due to the dynamics of repairing-and-decoding rules that can be potentially observed by eavesdroppers \cite{Sameer:ISIT2010}.
These drawbacks motivate the need for \emph{exact} repair of failed nodes. This leads to the following question: is it possible to achieve the cut-set lower bound region presented, with the extra constraint of exact repair?

Recently, significant progress has been made on the two extreme points of the family of Regenerating Codes (and arguably most interesting): the MBR point \cite{KVSKR:09} and the MSR point \cite{WuD:09,ShahRKR:09,SuhR:09}. The authors in \cite{KVSKR:09} showed that for $d=n-1$ (the interesting case), the optimal MBR point can be achieved with a deterministic scheme requiring a small finite-field size and repair bandwidth matching the cut-set bound of (\ref{MBR_cutset_d_n-1}).

For the MSR point, \cite{WuD:09} showed that it can be attained for the cases of $k=2$ and $k=n-1$ when $d=n-1$.
 Subsequently, the authors in \cite{ShahRKR:09} established that for $\frac{k}{n} > \frac{1}{2} + \frac{2}{n}$, cut-set bounds cannot be achieved for exact repair under \emph{scalar linear} codes (i.e., $\beta=1$) where symbols are not allowed to be split into arbitrarily small sub-symbols as with vector linear codes\footnote{This is equivalent to having large block-lengths in the classical setting. Under non-linear and vector linear codes, tightness of cut-set bounds remains open.}. For large $n$, this case boils down to $\frac{k}{n} > \frac{1}{2}$. For $\frac{k}{n} \leq \frac{1}{2}$, whether or not exact repair comes with a non-zero gap from cut-set bounds remained an open problem.

Recently, the authors in \cite{SuhR:09} showed that Exact-MSR codes can match the cut-set bound of (\ref{eq:MSRpoint}) for the case of $\frac{k}{n} \leq \frac{1}{2}$ and $d \geq 2k-1$.\footnote{The idea was inspired by the code structure in \cite{ShahRKR:09} where exact repair is guaranteed for the systematic part only. Indeed, it is shown in \cite{SuhR:09} that the code introduced in \cite{ShahRKR:09} for exact repair of only the systematic nodes can also be used to repair the non-systematic (parity) node failures exactly provided repair construction schemes are appropriately designed.} For the in-between regime $\frac{k}{n} \in (\frac{1}{2}, \frac{1}{2} + \frac{2}{n}]$, \cite{CullinaDH:09} and \cite{SuhR:09} showed that cut-set bounds are achievable for the case of $k=3$. For the most general Exact-MSR case, finding the fundamental limits in storage and repair bandwidth for all values of $(n,k,d)$ remains a challenging open problem. We now briefly summarize some of these recent results.

\subsection{Exact-MBR Codes}

\begin{theorem}[Exact-MBR Codes \cite{KVSKR:09}]
For $d=n-1$, the cutset lower bound of (\ref{MBR_cutset_d_n-1}) can be achieved with a deterministic scheme that requires a finite-field alphabet size of at most $\frac{(n-1)n}{2}$.
\end{theorem}

\begin{figure}
  \centering
  \includegraphics[width=8cm]{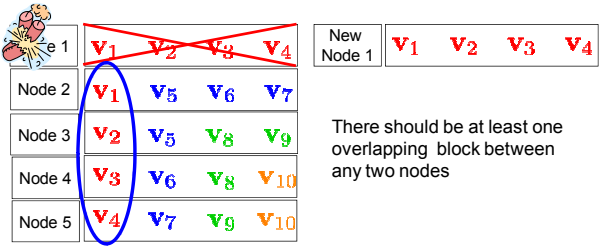}
  \caption{Repairing node 1 for a $(5,3)$-MBR code. Note that the number of desired blocks (that need to be repaired) is equal to the number of available equations (that can be downloaded). Hence, the code should be designed such that undesired blocks (interference) are totally avoided. \label{fig:53MBR}}
\end{figure}

Figure~\ref{fig:53MBR} illustrates an idea through the example of $(n,k,d,\alpha,\gamma)=(5,3,4,4,4)$ where the maximum file size of $\mathcal{M}=9$ (matching the cutset bound) can be stored. Let $\mathbf{a}$ be 9-dimensional data file. Each node stores 4 blocks with the form of $\mathbf{a}^t \mathbf{v}_i$, where $\mathbf{v}_i$ can be interpreted as a one-dimensional subspace of data file. We simply write only subspace vector to represent an actually stored block. Notice that the degree $d$ is equal to the number of storage blocks to be repaired, i.e., the number of available equations matches the number of desired variables for exact repair of a single node. Hence, for exact repair, there must be at least one duplicated block between node 1 and node $i$ for all $i \neq 1$.

This observation motivates the following idea. The idea is to have other nodes $i$ ($i \neq 1$) store each block of node 1, respectively: node 2, 3, 4, and 5 store $\mathbf{a}^t \mathbf{v}_1$, $\mathbf{a}^t \mathbf{v}_2$, $\mathbf{a}^t \mathbf{v}_3$, and $\mathbf{a}^t \mathbf{v}_4$ in its own place, respectively. Notice that for ensuring repair, it suffices to have only one duplicated block between any two storage nodes. Hence, node 2 can store another new 3 blocks of $\mathbf{a}^t \mathbf{v}_5$, $\mathbf{a}^t \mathbf{v}_6$ and $\mathbf{a}^t \mathbf{v}_7$ in the remaining other places. In accordance with the above procedure, node 3, 4, and 5 then copy each of three blocks in their space, respectively. We repeat this procedure until $10$ $(=4+3+2+1)$ blocks are stored in total. One can see that this construction guarantees exact repair of any failed node, since at least one block is duplicated between any two storage nodes and also the duplicated block is \emph{distinct}. See the example in Figure~\ref{fig:53MBR}.

The remaining issue is now to design these 10 subspace vectors $\mathbf{v}_i$, $i=1,\cdots,10$. The detailed construction comes from the MDS-code property that any three nodes out of five need to recover the whole data file. Observe in Figure~\ref{fig:53MBR} that nine distinct vectors can be downloaded from any three nodes. Hence, any $(10,9)$ MDS code can construct these $\mathbf{v}_i$'s. In this example, using the parity-check-code defined over ${\sf GF}(2)$, we can design the $\mathbf{v}_i$'s as follows: $\mathbf{v}_i = \mathbf{e}_i, \forall i =1,\cdots,9$ and $\mathbf{v}_{10} = [1,\cdots,1]^t$. It has been shown in~\cite{KVSKR:09} that this idea can be  extended to an arbitrary $(n,k)$ case.

This construction can be interpreted as an optimal \emph{interference avoidance} technique. To see this, observe in the figure that the number of desired blocks for exact repair matches the number of available equations that can be downloaded. Hence, the involvement of any undesired blocks (interference) precludes exact repair. A natural question arises: can this interference-avoidance technique provide solutions to the other extreme MSR point? It turns out that a new idea is needed to cover this point.

\subsection{Exact-MSR Codes}
The new idea is \emph{interference alignment} \cite{Mohammad:08,CadambeJ:08}.
The idea of interference alignment is to align multiple interference signals in a signal subspace whose dimension is smaller than the number of interferers. Specifically, consider the following setup where a decoder has to decode one desired signal which is linearly interfered with by two separate undesired signals. How many linear equations (relating to the number of channel uses) does the decoder need to recover its desired input signal? As the aggregate signal dimension spanned by desired and undesired signals is at most three, the decoder can naively recover its signal of interest with access to three linearly independent equations in the three unknown signals. However, as the decoder is interested in only one of the three signals, it can decode its desired unknown signal even if it has access to only two equations, provided the two undesired signals are judiciously aligned in a 1-dimensional subspace. See \cite{Mohammad:08,CadambeJ:08,SudTse:08} for details.

This concept relates intimately to our repair problem that involves recovery of a subset (related to the subspace spanned by a failed node) of the overall aggregate signal space (related to the entire user data dimension). This attribute was first observed in  \cite{WuD:09}, where it was shown that interference alignment could be exploited for Exact-MSR codes.

\begin{figure}
  \centering
  \includegraphics[width=9cm]{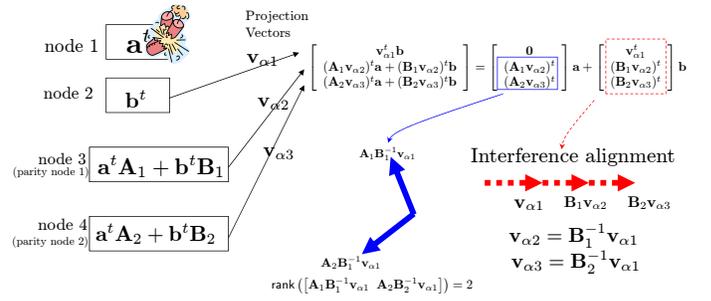}
  \caption{Repairing a $(4,2)$-MSR code, when node 1 fails \cite{WuD:09}.\label{fig:Repair42MDS}}
\end{figure}

Figure~\ref{fig:Repair42MDS} illustrates interference alignment for exact repair of failed node 1 for $(n,k,d,\alpha,\gamma)=(4,2,3,2,2)$ where the maximum file size of $\mathcal{M}=4$ can be stored. We introduce matrix notation for illustration purposes.
Let $\mathbf{a}= (a_1,a_2)^t$ and $\mathbf{b}= (b_1,b_2)^t$  be 2-dimensional information-unit vectors. Let $\mathbf{A}_i$ and $\mathbf{B}_i$ be $2$-by-$2$ encoding matrices for parity node $i$ ($i=1,2$), which contain encoding coefficients for the linear combination of $(a_1,a_2)$ and $(b_1,b_2)$, respectively. For example, parity node 1 stores blocks in the form of $\mathbf{a}^t \mathbf{A}_1  + \mathbf{b}^t \mathbf{B}_1$, as shown in Fig. \ref{fig:Repair42MDS}.
The encoding matrices for systematic nodes are not explicitly defined since those are trivially inferred.
Finally, we define 2-dimensional projection vectors $\mathbf{v}_{\alpha i}$'s ($i=1,2,3$) because of $\beta=1$.

Let us explain the interference alignment scheme.
First two blocks in each storage node are projected into a \emph{scalar} with projection vectors $\mathbf{v}_{\alpha i}$'s. By connecting to three nodes, we get: $ \mathbf{v}_{\alpha 1}^{t} \mathbf{b}$; $ (\mathbf{A}_1 \mathbf{v}_{\alpha 2})^t \mathbf{a} + (\mathbf{B}_1  \mathbf{v}_{\alpha 2})^t \mathbf{b}$;
$  ( \mathbf{A}_2 \mathbf{v}_{\alpha 3} )^t \mathbf{a} +  ( \mathbf{B}_2  \mathbf{v}_{\alpha 3} )^t \mathbf{b}$. Here the goal is to decode 2 desired unknowns out of 3 equations including 4 unknowns. To achieve this goal, we need:
\begin{align*}
{\sf rank} \left( \left[
    \begin{array}{c}
      (\mathbf{A}_1 \mathbf{v}_{\alpha 2})^t \\
      ( \mathbf{A}_2 \mathbf{v}_{\alpha 3} )^t \\
    \end{array}
  \right] \right) = 2; \;\; {\sf rank} \left( \left[
    \begin{array}{c}
      \mathbf{v}_{\alpha 1}^t \\
      (\mathbf{B}_1  \mathbf{v}_{\alpha 2})^t \\
      ( \mathbf{B}_2  \mathbf{v}_{\alpha 3} )^t \\
    \end{array}
  \right] \right) = 1.
\end{align*}
The second condition can be met by setting $\mathbf{v}_{\alpha 2} = \mathbf{B}_1^{-1} \mathbf{v}_{\alpha 1}$ and $\mathbf{v}_{\alpha 3} = \mathbf{B}_2^{-1} \mathbf{v}_{\alpha 1}$. This choice forces the interference space to be collapsed into a one-dimensional linear subspace, thereby achieving interference alignment.
On the other hand, we can satisfy the first condition as well by carefully choosing the $\mathbf{A}_i$'s and $\mathbf{B}_i$'s. For exact repair of node 2, we can apply the same idea. For parity node repair, we can remap parity node information and then apply the same technique.

It turned out this idea cannot be generalized to arbitrary $(n,k)$ case: it provides the optimal codes only for the case of $k=2$.
Recently, significant progress has been made: for the case of $\frac{k}{n} \leq \frac{1}{2}$, it has been shown that there is no price with exact repair for attaining the cutset lower bound of (\ref{eq:MSRpoint}).

\begin{theorem}[Exact-MSR Codes \cite{SuhR:09}] Suppose the MDS code rate is at most $\frac{1}{2}$, i.e., $\frac{k}{n} \leq \frac{1}{2}$ and the degree $d \geq 2k-1$. Then, the cutset bound of (\ref{eq:MSRpoint}) can be achieved with interference alignment. The achievable scheme is deterministic and requires a finite-field alphabet size of at most $2(n-k)$.
\end{theorem}

\begin{figure}
  \centering
  \includegraphics[width=9cm]{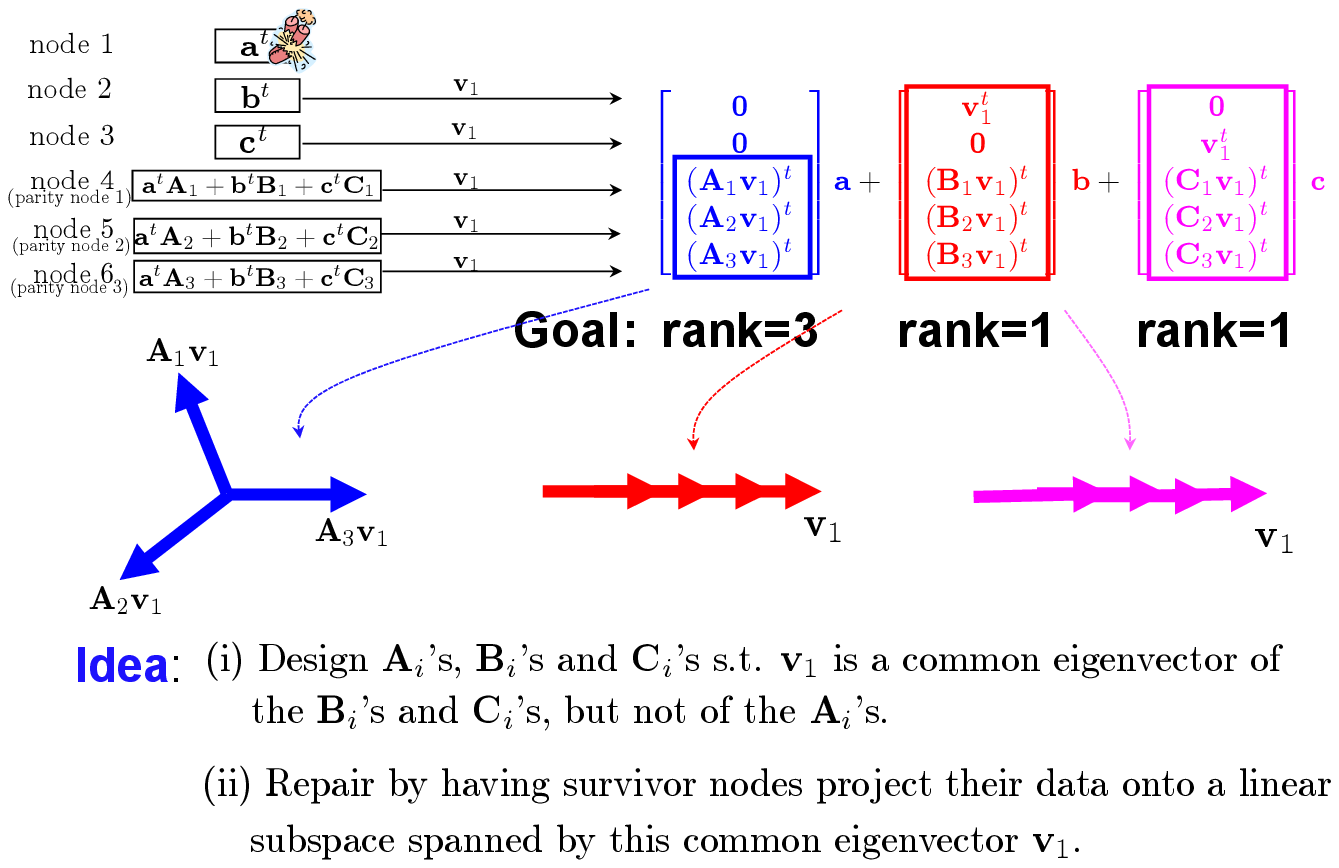}
  \caption{Repairing the $(6,3)$-MSR code when a systematic node fails. A common eigenvector concept is employed to achieve interference alignment simultaneously. \label{fig:63MSR}}
\end{figure}

A more sophisticated idea arises to cover this case: \emph{simultaneous interference alignment}. Figure~\ref{fig:63MSR} illustrates the interference alignment technique through the example of $(n,k,d,\alpha,\gamma) = (6,3,5,3,3)$ where $\mathcal{M}=9$. Let $\mathbf{a}= (a_1,a_2,a_3)^t$, $\mathbf{b}= (b_1,b_2,b_3)^t$ and $\mathbf{c}= (c_1,c_2,c_3)^t$ be 3-dimensional information-unit vectors. Let $\mathbf{A}_i$, $\mathbf{B}_i$ and $\mathbf{C}_i$ be $3$-by-$3$ encoding matrices for parity node $i$ ($i=1,2,3$). We define 3-dimensional projection vectors $\mathbf{v}_{\alpha i}$'s ($i=1,\cdots,5$).

By connecting to five nodes, we get five equations shown in the figure.
In order to successfully recover the desired signal components of $\mathbf{a}$, the matrix associated with $\mathbf{a}$ should have full rank of 3, while the other matrices corresponding to $\mathbf{b}$ and $\mathbf{c}$ should have rank 1, respectively.
In accordance with the $(4,2)$ code example in Figure~\ref{fig:Repair42MDS}, if one were to set $\mathbf{v}_{\alpha 3} = \mathbf{B}_1^{-1} \mathbf{v}_{\alpha 1}$, $\mathbf{v}_{\alpha 4} = \mathbf{B}_2^{-1} \mathbf{v}_{\alpha 2}$ and $\mathbf{v}_{\alpha 5} = \mathbf{B}_3^{-1} \mathbf{v}_{\alpha 1}$, then it is possible to achieve interference alignment with respect to $\mathbf{b}$. However, this choice also specifies the interference space of $\mathbf{c}$. If the $\mathbf{B}_i$'s and $\mathbf{C}_i$'s are not designed judiciously, interference alignment is not guaranteed for $\mathbf{c}$. Hence, it is not evident how to achieve interference alignment \emph{at the same time}.

In order to address the challenge of simultaneous interference alignment, a \emph{common eigenvector} concept is invoked. The idea consists of two parts: (i) designing the $(\mathbf{A}_i,\mathbf{B}_i, \mathbf{C}_i)$'s such that $\mathbf{v}_1$ is a common eigenvector of the $\mathbf{B}_i$'s and $\mathbf{C}_i$'s, but not of $\mathbf{A}_i$'s\footnote{Of course, five additional constraints also need to be satisfied for the other five failure configurations for this $(6,3,5)$ code example.}; (ii) repairing by having survivor nodes \emph{project} their data onto a linear subspace spanned by this common eigenvector $\mathbf{v}_1$.
We can then achieve interference alignment for $\mathbf{b}$ and $\mathbf{c}$ at the same time, by setting $\mathbf{v}_{\alpha i} = \mathbf{v}_1, \forall i$. As long as $[\mathbf{A}_1 \mathbf{v}_1, \mathbf{A}_2 \mathbf{v}_1, \mathbf{A}_3 \mathbf{v}_1 ]$ is invertible, we can also guarantee the decodability of $\mathbf{a}$. See Figure~\ref{fig:63MSR}.

\begin{figure*}[t]
\begin{center}
{\epsfig{figure=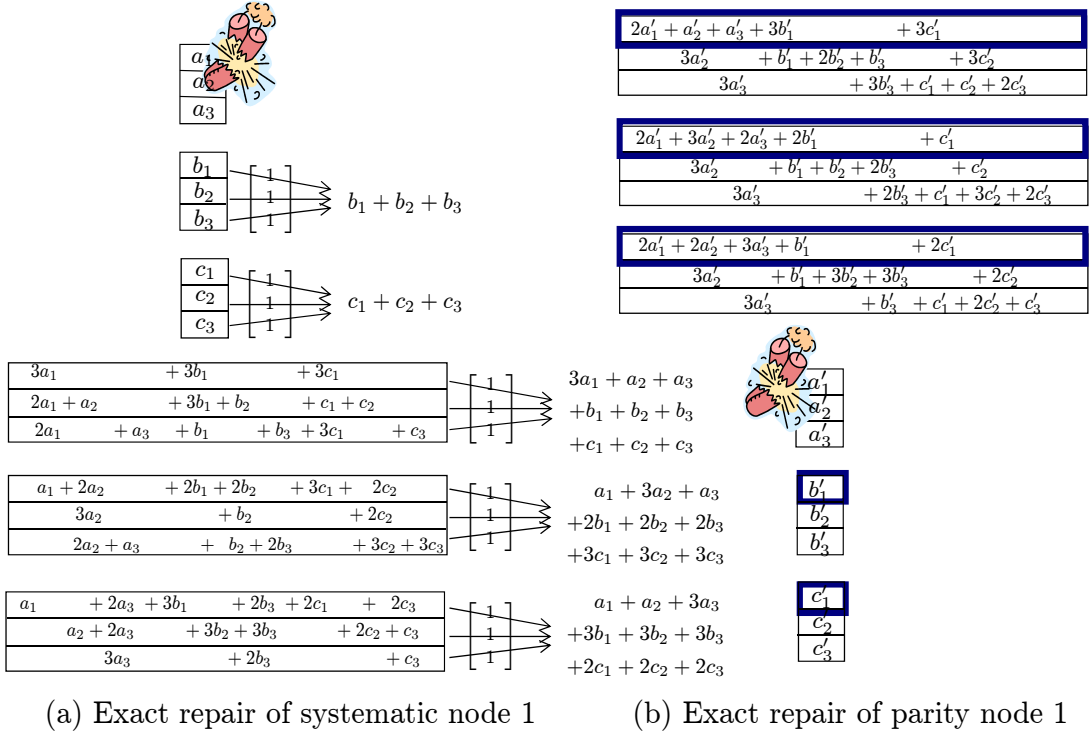, angle=0, width=0.8\textwidth}}
\end{center}
\caption{Illustration of exact repair for a $(6,3,5)$ E-MSR code defined over ${\sf GF}(4)$ where a generator polynomial $g(x)= x^2 + x +1$. The solution for systematic node repair is simple: setting all of the projection vectors as $(1,1,1)^t$. This enables simultaneous interference alignment, while guaranteeing the decodability of $\mathbf{a}$. For our carefully chosen parameters, parity node repair is much simpler. For the repair, we download only the first equation from each survivor node to solve five linear equations containing only five unknowns. } \label{fig:63_EMSR_Orthogonal}
\end{figure*}

The challenge is now to design encoding matrices to guarantee the existence of a common eigenvector while also satisfying the decodability of desired signals. The difficulty comes from the fact that in the $(6,3,5)$ code example, these constraints need to be satisfied for \emph{all} six possible failure configurations. The structure of \emph{elementary matrices} (generalized matrices of Householder and Gauss matrices) gives insights into this. To see this, consider a 3-by-3 elementary matrix $\mathbf{A}$:
\begin{align}
\mathbf{A} =  \mathbf{u} \mathbf{v}^t + \alpha \mathbf{I},
\end{align}
where $\mathbf{u}$ and $\mathbf{v}$ are 3-dimensional vectors. Note that the dimension of the null space of $\mathbf{v}$ is 2 and the null vector $\mathbf{v}^\bot$ is an eigenvector of $\mathbf{A}$, i.e., $\mathbf{A} \mathbf{v}^{\bot} = \alpha \mathbf{v}^{\bot}$.
This motivates the following structure:
\begin{align}
\begin{split}
\label{eq:63_EncodingMatrices}
\mathbf{A}_1 &= \mathbf{u}_1 \mathbf{v}_1^t  + \alpha_1 \mathbf{I}; \; \mathbf{B}_1 = \mathbf{u}_1 \mathbf{v}_2^t  + \beta_1 \mathbf{I}; \;
\mathbf{C}_1 = \mathbf{u}_1 \mathbf{v}_3^t  + \gamma_1 \mathbf{I} \\
\mathbf{A}_2 &= \mathbf{u}_2 \mathbf{v}_1^t  + \alpha_2 \mathbf{I}; \; \mathbf{B}_2 = \mathbf{u}_2 \mathbf{v}_2^t + \beta_2 \mathbf{I}; \;
\mathbf{C}_2 = \mathbf{u}_2 \mathbf{v}_3^t+ \gamma_2 \mathbf{I} \\
\mathbf{A}_3 &= \mathbf{u}_3 \mathbf{v}_1^t  + \alpha_3 \mathbf{I}; \;
\mathbf{B}_3 = \mathbf{u}_3 \mathbf{v}_2^t + \beta_3 \mathbf{I}; \;
\mathbf{C}_3 =\mathbf{u}_3 \mathbf{v}_3^t + \gamma_3 \mathbf{I},
\end{split}
\end{align}
where $\mathbf{v}_i$'s are 3-dimensional linearly independent vectors and so are $\mathbf{u}_i$'s. The values of the $\alpha_i$'s, $\beta_i$'s and $\gamma_i$'s can be arbitrary non-zero values. For simplicity, we consider the simple case where the $\mathbf{v}_i$'s are \emph{orthonormal}, although these need not be orthogonal, but only linearly independent. We then see that $\forall i=1,2,3,$
\begin{align}
\begin{split}
\mathbf{A}_i \mathbf{v}_1 &=  \alpha_i \mathbf{v}_1 + \mathbf{u}_i,\\
\mathbf{B}_i \mathbf{v}_1 &= \beta_i \mathbf{v}_1,\\
\mathbf{C}_i \mathbf{v}_1 &= \gamma_i \mathbf{v}_1.
\end{split}
\end{align}
Importantly, notice that $\mathbf{v}_1$ is a common eigenvector of the $\mathbf{B}_i$'s and $\mathbf{C}_i$'s, while simultaneously ensuring that the vectors of $\mathbf{A}_i \mathbf{v}_1$ are linearly independent. Hence, setting $\mathbf{v}_{\alpha i} = \mathbf{v}_1$ for all $i$, it is possible to achieve simultaneous interference alignment while also guaranteeing the decodability of the desired signals. On the other hand, this structure also guarantees exact repair for $\mathbf{b}$ and $\mathbf{c}$. We use $\mathbf{v}_2$ for exact repair of $\mathbf{b}$. It is a common eigenvector of the $\mathbf{C}_i$'s and $\mathbf{A}_i$'s, while ensuring $[\mathbf{B}_1 \mathbf{v}_2, \mathbf{B}_2 \mathbf{v}_2, \mathbf{B}_3 \mathbf{v}_2]$ invertible. Similarly, $\mathbf{v}_3$ is used for $\mathbf{c}$.

Parity nodes can be repaired by drawing a \emph{dual} relationship with systematic nodes.
The procedure has two steps. The first is to remap parity nodes with
$\mathbf{a}'$, $\mathbf{b}'$, and $\mathbf{c}'$, respectively.
Systematic nodes can then be rewritten in terms of the prime notations:
\begin{align}
\begin{split}
\mathbf{a}^t &= \mathbf{a}'^t \mathbf{A}_1' + \mathbf{b}'^t \mathbf{B}_1' +
\mathbf{c}'^t \mathbf{C}_1', \\
\mathbf{b}^t &= \mathbf{a}'^t \mathbf{A}_2' + \mathbf{b}'^t \mathbf{B}_2' +
\mathbf{c}'^t \mathbf{C}_2', \\
\mathbf{c}^t &= \mathbf{a}'^t \mathbf{A}_3' + \mathbf{b}'^t \mathbf{B}_3' +
\mathbf{c}'^t \mathbf{C}_3',
\end{split}
\end{align}
where the newly mapped encoding matrices $(\mathbf{A}_i', \mathbf{B}_i', \mathbf{C}_i)$'s are defined as:
\begin{align}
\label{eq-63_mapping}
\left[
   \begin{array}{ccc}
     \mathbf{A}_1'  & \mathbf{A}_2' & \mathbf{A}_3' \\
     \mathbf{B}_1'  & \mathbf{B}_2' & \mathbf{B}_3' \\
\mathbf{C}_1' & \mathbf{C}_2' & \mathbf{C}_3' \\
   \end{array}
 \right]: = \left[
   \begin{array}{ccc}
     \mathbf{A}_1  & \mathbf{A}_2 & \mathbf{A}_3 \\
     \mathbf{B}_1  & \mathbf{B}_2 & \mathbf{B}_3 \\
\mathbf{C}_1 & \mathbf{C}_2 & \mathbf{C}_3 \\
   \end{array}
 \right]^{-1}.
\end{align}
With this remapping, one can dualize the relationship between systematic and parity node repair. Specifically, if all of the $\mathbf{A}_i'$'s, $\mathbf{B}_i'$'s, and $\mathbf{C}_i'$'s are \emph{elementary matrices} and form a similar code-structure as in (\ref{eq:63_EncodingMatrices}), exact repair of the parity nodes becomes transparent.
It was shown that a special relationship between $[\mathbf{u}_1, \mathbf{u}_2, \mathbf{u}_3]$ and $[\mathbf{v}_1, \mathbf{v}_2, \mathbf{v}_3]$ through the correct choice of ($\alpha_i$, $\beta_i$, $\gamma_i$)'s can also guarantee the \emph{dual} structure of (\ref{eq:63_EncodingMatrices})~\cite{SuhR:09}.

Figure~\ref{fig:63_EMSR_Orthogonal} shows a numerical example for exact repair of $(a)$ systematic node 1 and $(b)$ parity node 1 where $[\mathbf{v}_1, \mathbf{v}_2, \mathbf{v}_3]=[2, 2, 2; 2,3,1; 2,1,3]$. This example illustrates the code structure that generalizes the code introduced in~\cite{ShahRKR:09}. See~\cite{SuhR:09} for details. This generalized code structure allows for a much larger design space for exact repair.

Notice that the projection vector solution for systematic node repair is simple: $\mathbf{v}_{\alpha i}= 2^{-1}\mathbf{v}_1=(1,1,1)^t, \forall i$.
Note that this choice enables simultaneous interference alignment, while guaranteeing the decodability of $\mathbf{a}$. Notice that $(b_1,b_2,b_3)$ and  $(c_1,c_2,c_3)$ are aligned into $b_1+b_2+b_3$ and $c_1+c_2+c_3$, respectively, while three equations associated with $\mathbf{a}$ are linearly independent.

The dual structure also guarantees exact repair of parity nodes. Importantly, we have chosen code parameters from the generalized code structure of~\cite{SuhR:09} such that parity node repair is quite simple. As shown in Figure~\ref{fig:63_EMSR_Orthogonal} $(b)$, downloading only the first equation from each survivor node ensures exact repair. Notice that the five downloaded equations contain only five unknown variables of $(a_1',a_2',a_3', b_1',c_1')$ and three equations associated with $\mathbf{a}'$ are linearly independent. Hence, we can successfully recover $\mathbf{a}'$.

It has been shown in \cite{SuhR:09} that this alignment technique can be easily generalized to arbitrary $(n,k,d)$ where $n \geq 2k$ and $d \geq 2k-1$.

\section{Model III: Exact Repair of the Systematic Part}
\label{sec:hybrid}

In this section, we review the constructive scheme given in \cite{Wu:09c}, which gives a construction of systematic $(n,k)$-MDS codes for $2k\le n$ that achieves the minimum repair bandwidth when repairing from $k+1$ nodes.

\begin{figure}
  \centering
  \includegraphics[width=8cm]{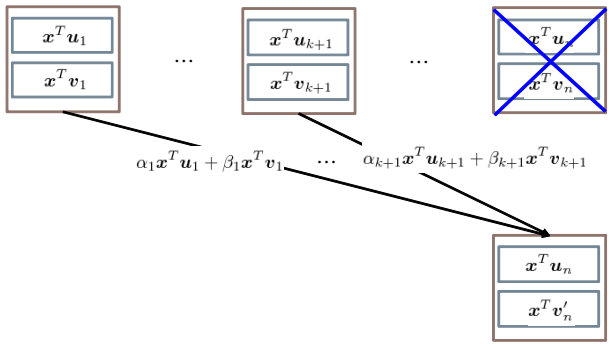}\\
  \caption{Illustration of the scheme in \cite{Wu:09c}.}\label{fig:hybrid}
\end{figure}
The scheme is illustrated in Figure~\ref{fig:hybrid}. Let $\mathbb{F}$ denote the finite field where the code is defined in. In Figure~\ref{fig:hybrid}, $\boldsymbol{x}\in \mathbb{F}^{2k}$ is a vector consisting of the  $2k$ original information symbols. Each node stores 2 symbols, $\boldsymbol{x}^T \boldsymbol{u}_i$ and $\boldsymbol{x}^T\boldsymbol{v}_i$. The vectors $\{\boldsymbol{u}_i\}$ do not change over time but  $\{\boldsymbol{v}_i\}$ change as the code repairs. We maintain the invariant property that the $2n$ length-$2k$ vectors $\{\boldsymbol{u}_i,\boldsymbol{v}_i\}$ form an $(2n,2k)$-MDS code; that is, any $2k$ vectors in the set $\{\boldsymbol{u}_i,\boldsymbol{v}_i\}$ have full rank $2k$. This certainly implies that the $n$ nodes form an $(n,k)$-MDS code. We initialize the code using any $(2n,2k)$ systematic MDS code over $\mathbb{F}$.

Now we consider the situation of a repair. Without loss of generality, suppose node $n$ failed and is repaired by accessing nodes $1,\ldots,k+1$. As illustrated in Figure~\ref{fig:hybrid}, the replacement node downloads $\alpha_i\boldsymbol{x}^T\boldsymbol{u}_i+\beta_i\boldsymbol{x}^T\boldsymbol{v}_i$ from each node of $\{1,\ldots,k+1\}$. Using these $k+1$ downloaded symbols, the replacement node computes two symbols $\boldsymbol{x}^T \boldsymbol{u}_n$ and $\boldsymbol{x}^T\boldsymbol{v}'_n$ as follows:
\begin{align}
\sum_{i=1}^{k+1} \left(\alpha_i\boldsymbol{x}^T\boldsymbol{u}_i+\beta_i\boldsymbol{x}^T\boldsymbol{v}_i\right)=\boldsymbol{x}^T\boldsymbol{u}_n\label{eq:1}\\
\sum_{i=1}^{k+1} \rho_i \left(\alpha_i\boldsymbol{x}^T\boldsymbol{u}_i+\beta_i\boldsymbol{x}^T\boldsymbol{v}_i\right)=\boldsymbol{x}^T\boldsymbol{v}'_n\label{eq:2}
\end{align}
Note that $\boldsymbol{v}'_n$ is allowed to be different from  $\boldsymbol{v}_n$; the property that we maintain is that the repaired code continues to be an $(2n,2k)$-MDS code. Here $\{\alpha_i,\beta_i,\rho_i\}$ and $\boldsymbol{v}'_n$ are the variables that we can control. The following theorem shows that we can choose these variables so that (\ref{eq:1}) and (\ref{eq:2}) are satisfied and the repaired code continues to be an $(2n,2k)$-MDS  code.

\begin{theorem}[\cite{Wu:09c}]~\\
Let $\mathbb{F}$ be a finite field whose size is greater than
\begin{align}
d_0 = 2\left(\begin{array}{c}
               2n-1 \\
               2k-1
             \end{array}
\right).
\end{align}
 Suppose the old code specified by $\{\boldsymbol{u}_i,\boldsymbol{v}_i\}$ is an $(2n,2k)$-MDS code defined over $\mathbb{F}$. When node $n$ fails, there exists an assignment of the variables $\{\alpha_i,\beta_i,\rho_i\}$ such that (\ref{eq:1}) and (\ref{eq:2}) are satisfied and the repaired code continues to be an $(2n,2k)$-MDS  code.
 \label{thrm:existence}
\end{theorem}

\begin{corollary}[A Systematic $(n,k)$-MDS Code]~\\
The above scheme gives a construction of systematic $(n,k)$-MDS codes for $2k\le n$ that achieves the minimum repair bandwidth when repairing from $k+1$ nodes.
\end{corollary}
{\bf Proof:~} Consider $n\ge 2k$. Note that in the above scheme, we can initialize the code $\{\boldsymbol{u}_1,\ldots,\boldsymbol{u}_n, \boldsymbol{v}_1,\ldots,\boldsymbol{v}_n\}$ with any $(2n,2k)$-MDS code. In particular, we can use a systematic code and assign the $2k$ systematic code vectors to $\{\boldsymbol{u}_1,\ldots,\boldsymbol{u}_{2k}\}$. Since $\{\boldsymbol{u}_1,\ldots,\boldsymbol{u}_n\}$ do not change over time, the code remains a systematic $(2n,2k)$-MDS code. Thus the $n$ nodes form a systematic $(n,k)$-MDS code. The code repairs a failure by downloading $k+1$ blocks from $d=k+1$ nodes, with the total file size is $\size=2k$, achieving the cut-set bounds derived
in section~\ref{sec:functional}. \hfill \qed

\section{Discussion and Conclusions}

We provided an overview of recent results about the problem of reducing repair traffic in distributed storage systems based on erasure coding. Three versions of the repair problems are considered: \emph{exact repair}, \emph{functional repair} and \emph{exact repair of systematic parts}. In the exact repair model, the lost content is exactly regenerated; in the functional repair model, only the same MDS-code property is maintained before and after repairing; in the exact repair of systematic parts, the systematic part is exactly reconstructed but the non-systematic part follows a functional repair model.

The functional repair problem is in essence a problem of multicasting from a source to an unbounded number of receivers over an unbounded graph. As we showed there is a tradeoff between storage and repair bandwidth
and the two extremal points are achieved by Minimum Bandwidth and Minimum Storage regenerating (MBR and MSR) codes.
The repair bandwidth is characterized by the min-cut bounds and
therefore the functional repair problem is completely solved.

\begin{table}[t]
\caption{Known results for exact MBR and MSR codes. All points correspond to regimes where the cut-set bound region is known to be achievable}
\centering
\begin{tabular}{|c|l|l|}
\hline
   & \qquad \quad MBR & \qquad \quad MSR \\
\multirow{4}{*}{Functional} &  &  \\
\hline
& \multicolumn{2}{|c|}{ } \\
& \multicolumn{2}{|c|}{ \cite{DimakisGWWR:08}: $\forall n,k,d$  }\\
& \multicolumn{2}{|c|}{ } \\  \hline
\multirow{4}{*}{Hybrid} &  &  \\
&  \qquad \quad ? &   \cite{ShahRKR:09}: $\frac{k}{n} \leq \frac{1}{2}$, $d \geq 2k-1$ \\
&   &   \\
&  & \cite{Wu:09c}, \cite{KVSKR:09}: $\frac{k}{n}\leq \frac{1}{2}$, $d=k+1$  \\
&  &   \\ \hline
\multirow{4}{*}{Exact} &  &  \\
& \cite{KVSKR:09}:$d=n-1$  & \cite{SuhR:09}:$\frac{k}{n}\leq \frac{1}{2}$, $d\geq 2k-1$\\
&  &   \\ \hline
\end{tabular}
\end{table}

Problems that require exact repair correspond to network coding problems having sinks with overlapping subset demands. For such problems
cut-set bounds are not tight in general and linear codes might not even suffice~\cite{DFZ05}.
The recent work we discussed~\cite{KVSKR:09} showed that for MBR codes the repair bandwidth given by the cut-set bound is achievable for the interesting case of $d=n-1$. The minimum-storage point seems harder to understand. The best known constructions~\cite{SuhR:09} we
presented match the cut-set bound for $k/n\leq 1/2$
for the interesting regime of connectivity $d \in [2k-1,n-1]$.
A corresponding negative result~\cite{ShahRKR:09} established that for $\frac{k}{n} > \frac{1}{2} + \frac{2}{n}$, the cut-set bound cannot be achieved by interference alignment-based linear schemes.
Table I summarizes what is known for the repair bandwidth region and an online editable bibliography (wiki) can be found online~\cite{storagewiki}. All the cases marked correspond to 
regimes where the cut-set bound is known to be achievable. To the best of our knowledge there 
are no information theoretic upper bounds other than the cut-set bound and it would be
very interesting to see if the region could be universally achievable. Of particular interest
is the case of exact Minimum Storage Regenerating codes for $d=n-1$ and high rates.

In addition to the complete characterization of the repair rate region for storage, there are several other interesting open problems.
A first problem is to investigate the influence of network topology, as initiated  recently~\cite{LiInfocom10} for trees. All the prior work so far has been assuming a complete connectivity topology for the storage network. However, most networks of interest will have different communication capacities and sparse topologies. For these cases communication will 
have a different cost and it would be interesting to formulate this as an optimization problem. 

Secondly, the issues of security and privacy are important for distributed storage. When coding is used, errors can be propagated in several mixed blocks through the repair process ~\cite{Dikaliotis:ISIT2010} and an error-control mechanism is required. A related issue is that of privacy of the data by information leakage to eavesdroppers during repairs\cite{Sameer:ISIT2010}.

Finally small finite-field constructions require further investigation. While many of the constructions presented require a large finite-field size, practical storage systems would benefit from efficient binary operations. Recently Zhang et al. suggested a scheme for repairing Evenodd codes~\cite{ITA_arrayrepair}, which are binary codes with $n=k+2$. While the proposed scheme does not match the cut-set bound it improves on the naive repairing method of reconstructing all the data blocks. Constructing regenerating codes for small finite fields or designing repair algorithms for existing codes will be of significant practical interest.

\section{Acknowledgments}
We gratefully acknowledge Prof. P. V. Kumar (of IISs) and his students, N. B. Shah and K. V. Rashmi, for insightful discussions and fruitful collaboration.

\bibliographystyle{IEEEtran}

\footnotesize

\begin{thebibliography}{1}
\providecommand{\url}[1]{#1}
\csname url@samestyle\endcsname
\providecommand{\newblock}{\relax}
\providecommand{\bibinfo}[2]{#2}
\providecommand{\BIBentrySTDinterwordspacing}{\spaceskip=0pt\relax}
\providecommand{\BIBentryALTinterwordstretchfactor}{4}
\providecommand{\BIBentryALTinterwordspacing}{\spaceskip=\fontdimen2\font plus
\BIBentryALTinterwordstretchfactor\fontdimen3\font minus
  \fontdimen4\font\relax}
\providecommand{\BIBforeignlanguage}[2]{{%
\expandafter\ifx\csname l@#1\endcsname\relax
\typeout{** WARNING: IEEEtran.bst: No hyphenation pattern has been}%
\typeout{** loaded for the language `#1'. Using the pattern for}%
\typeout{** the default language instead.}%
\else
\language=\csname l@#1\endcsname
\fi
#2}}
\providecommand{\BIBdecl}{\relax}
\BIBdecl


\bibitem{storagewiki}
The Coding for Distributed Storage wiki 
\texttt{http://tinyurl.com/storagecoding}

\bibitem{WeatherspoonK:02}
H.~Weatherspoon and J.~D. Kubiatowicz, ``Erasure coding vs. replication: a
  quantitiative comparison,'' in {\em Proc. IPTPS}, 2002.

\bibitem{Kubiatowicz+:00}
J. Kubiatowicz, D. Bindel, Y. Chen, S. Czerwinski, P. Eaton, D. Geels, R. Gummadi, S. Rhea, H. Weatherspoon, W. Weimer, C. Wells, and B. Zhao, ``OceanStore: An architecture for global-scale persistent storage," in \emph{Proceedings of the Ninth International Conference on Architectural Support for Programming Languages and Operating Systems (ASPLOS)}, Boston, MA, Nov. 2000.

\bibitem{RheaWEG+:01}
S. Rhea, C. Wells, P. Eaton, D. Geels, B. Zhao, H. Weatherspoon, and
J. Kubiatowicz, ``Maintenance-free global data storage,'' \emph{IEEE Internet
Computing}, pp. 40-49, September 2001.

\bibitem{bhagwan04total}
R.~Bhagwan, K.~Tati, Y.-C. Cheng, S.~Savage, and G.~M. Voelker, ``Total recall:
  System support for automated availability management,'' in {\em NSDI}, 2004.



\bibitem{ReedSolomon}
I.~Reed and G.~Solomon, ``Polynomial codes over certain finite fields,'' in
{\em Journal of the SIAM}, 1960.

\bibitem{Rabin}
M.~O. Rabin, ``Efficient dispersal of information for security, load balancing
  and fault tolerance,'' in {\em Journal of the ACM, 36(2):335--348}, 1989.


\bibitem{Raptor}
A.~Shokrollahi, ``Raptor codes,'' {\em IEEE Trans. on Information Theory}, June
  2006.

\bibitem{mct}
T.~Richardson and R.~Urbanke, {\em Modern {C}oding {T}heory}.
\newblock Cambridge University Press, 2008.



\bibitem{evenodd}
M.~Blaum, J.~Brady, J.~Bruck, and J.~Menon, ``{EVENODD}: An efficient scheme
  for tolerating double disk failures in raid architectures,'' in {\em IEEE
  Transactions on Computers}, 1995.

\bibitem{array1}
M.~Blaum, J.~Bruck, and A.~Vardy, ``{MDS} array codes with independent parity
  symbols,'' in {\em IEEE Transactions o Information Theory}, 1996.

\bibitem{array2}
M.~Blaum, P.~G. Farrell, and H.~van Tilborg, ``Book chapter on array codes,''
  in {\em Handbook of Coding Theory, V. S. Pless and W. C. Huffman, Eds.},
  1998.

\bibitem{SiegelBook}
B.~Marcus, R.~M. Roth, and P.~Siegel, ``Constrained systems and coding for
  recording channels,'' {\em Handbook of Coding Theory, V. Pless and W.C.
  Huffman (editors)}, pp.~1635--1764, 1998.

\bibitem{ITA_arrayrepair}
Z.~Wang, R.~Mateescu, A.G.~Dimakis, J.~Bruck,
``Array codes for distributed storage: Results and open problems'',
Information Theory and Applications (ITA), 2010.


\bibitem{Ahlswede00}
L.~Ahlswede, N.~Cai, S.-Y.~R. Li, and R.~W. Yeung, ``Network information
  flow,'' {\em IEEE Trans. Info. Theory}, vol.~46, no.~4, pp.~1204--1216, 2000.


\bibitem{LYC03}
S.-Y.~R. Li, R.~W. Yeung, and N.~Cai, ``Linear network coding,'' \emph{IEEE
  Trans. on Information Theory}, vol.~49, pp. 371--381, February 2003.

\bibitem{lima2}
L.~Lima, J.~Barros, M.~Medard, and A.~Toledo, ``Protecting the code: Secure
  multiresolution network coding,'' in {\em IEEE Information Theory Workshop
  (ITW)}, 2009.

\bibitem{KM03}
R.~Koetter and M.~M\'edard, ``An algebraic approach to network coding,'' {\em
  Transactions on Networking}, October 2003.

\bibitem{RLNC}
T.~Ho, M.~M\'{e}dard, R.~Koetter, D.~Karger, M.~Effros, J.~Shi, and B.~Leong,
  ``A random linear network coding approach to multicast,'' {\em IEEE
  Transactions on Information Theory}, October 2006.

\bibitem{JaggiSCEEJT:05}
S.~Jaggi, P.~Sanders, P.~A. Chou, M.~Effros, S.~Egner, K.~Jain, and
  L.~Tolhuizen, ``Polynomial time algorithms for network code construction,''
  \emph{IEEE Trans.\ Inform.\ Theory}, vol.~51, pp. 1973--1982, Jun. 2005.

\bibitem{Fragouli_primer}
C.~Fragouli, J.~L. Boudec, and J.~Widmer, ``Network coding: an instant
  primer,'' {\em ACM SIGCOMM Computer Comm. Review}, 2006.

\bibitem{DFZ05}
R.~Dougherty, C.~Freiling, and K.~Zeger, ``Insufficiency of linear coding in
  network information flow,'' {\em IEEE Transactions on Information Theory},
  August 2005.

\bibitem{Jiang06}
A.~Jiang, ``Network coding for joint storage and transmission with minimum
  cost,'' in {\em International Symposium on Information Theory (ISIT)}, July
  2006.






\bibitem{DimakisGWWR:08}
A.~G. Dimakis, P.~G. Godfrey, Y. Wu, M.~J. Wainwright, and K.~Ramchandran, ``Network coding
for distributed storage systems," IEEE Transactions on Information Theory, to appear.

\bibitem{WuDR:07}
Y.~Wu, A.~G. Dimakis, and K.~Ramchandran, ``Deterministic regenerating codes for
  distributed storage,'' in \emph{Allerton Conference on Control, Computing,
  and Communication}, Monticello, IL, October 2007.

\bibitem{Wu:09}
Y.~Wu, ``Existence and construction of capacity-achieving network codes for
  distributed storage,'' in \emph{Proc.\ IEEE Int'l Symp.\ Information
  Theory}, Seoul, Korea, June 2009.

\bibitem{Wu:10}
Y.~Wu.
\newblock ``Existence and construction of capacity-achieving network codes for distributed storage,''
\newblock In {\em IEEE Journal on Selected Areas in Communications} (JSAC), vol. 28, issue 2, pp. 277--288, Feb. 2010.

\bibitem{Mohammad:08}
M. A. Maddah-Ali and S. A. Motahari and A. K. Khandani, ``Communication over {MIMO} {X} Channels: Interference Alignment, Decomposition, and Performance Analysis," \emph{IEEE Transactions on Information Theory}, 54(8), pp. 3457--3470, Aug. 2008.

\bibitem{CadambeJ:08}
V. R. Cadambe and  S. A. Jafar, ``Interference alignment and the degrees of freedom for the $K$ user interference channel," \emph{IEEE Transactions on Information Theory}, 54(8), pp. 3425--3441, Aug. 2008.

\bibitem{SudTse:08}
C. Suh and  D. Tse, ``Interference alignment for cellular networks,"
\newblock In {\em Allerton Conference on Control, Computing, and Communication}, Urbana-Champaign, IL, Sep. 2008.

\bibitem{WuD:09}
Y. Wu and A.~G. Dimakis, ``Reducing repair traffic for erasure coding-based storage via interference alignment," in \emph{Proc.\ IEEE Int. Symp. on Information Theory (ISIT)}, Seoul, Korea, July 2009.


\bibitem{CullinaDH:09}
D. Cullina, A.~G. Dimakis, and T. Ho, ``Searching for minimum storage regenerating codes,"
\newblock In {\em Allerton Conference on Control, Computing, and
  Communication}, Urbana-Champaign, IL, September 2009.

\bibitem{KVSKR:09}
Rashmi K.V., N. B. Shah, P. V. Kumar, and K. Ramchandran
``Exact regenerating codes for distributed storage," In {\em Allerton Conference on Control, Computing, and
  Communication}, Urbana-Champaign, IL, September 2009 (preprint available at http://arxiv.org/abs/0906.4913).

\bibitem{ShahRKR:09}
N. B. Shah, K. V. Rashmi, P. V. Kumar, and K. Ramchandran, ``Explicit codes minimizing repair bandwidth for distributed storage," in \emph{Proc.\ IEEE ITW}, Jan. 2010 (preprint available at http://arxiv.org/abs/0908.2984).

\bibitem{SuhR:09}
C. Suh and K. Ramchandran, ``Exact Regeneration Codes for Distributed Storage Repair Using Interference Alignment," to appear in \emph{Proc.\ IEEE Int. Symp. on Information Theory (ISIT)}, June 2010 (Preprint available at
http://arxiv.org/abs/1001.0107v2).

\bibitem{Wu:09c}
Y. Wu.
\newblock ``A construction of systematic MDS codes with minimum repair bandwidth,"
\newblock Submitted to \emph{IEEE Transactions on Information Theory}, Aug. 2009. Preprint available at
http://arxiv.org/abs/0910.2486.

\bibitem{Dum09}
A. Duminuco and E.Biersack,
``A practical study of regenerating codes for peer-to-peer backup systems,''
Proceedings of the International Conference on Distributed Computing Systems (ICDCS) 2009.

\bibitem{LiInfocom10}
J. Li and S. Yang and Xin Wang and Baochun Li.
``Tree-structured Data Regeneration in Distributed Storage Systems with Regenerating Codes''
Proceedings of IEEE INFOCOM 2010.

\bibitem{Dikaliotis:ISIT2010}
T. Dikaliotis and A.~G. Dimakis and T. Ho
``Security in Distributed Storage Systems by Communicating a Logarithmic Number of Bits''
\emph{submitted to IEEE ISIT}, July 2010.

\bibitem{Sameer:ISIT2010}
S. Pawar and S. El Rouayheb and K. Ramchandran, ``On Security for Distributed Storage Systems,"
\emph{submitted to IEEE ISIT}, July 2010.


\end{thebibliography}

\begin{biography}{Alexandros G. Dimakis} is an assistant Professor at the Viterbi School of Engineering at the University of Southern California. He has been a faculty member in the Department of Electrical Engineering - Systems since 2009. He received his Ph.D. in 2008 and M.S. degree in 2005 in electrical engineering and computer science, both from the University of California, Berkeley. Prior to that, he obtained the Diploma degree in Electrical and Computer Engineering from the National Technical University of Athens in 2003.

He received the Eliahu Jury award in 2008 for his thesis work on codes for distributed storage, two outstanding paper awards, the UC Berkeley Regents Fellowship and the Microsoft Research Fellowship. He was a postdoctoral scholar at the Center for the Mathematics of Information (CMI) at Caltech in 2008. His research interests include communications, coding theory, signal processing, and networking, with a current focus on distributed storage, network coding, large-scale inference and message passing algorithms.
\end{biography}

\begin{biography}{Kannan Ramchandran} is a Professor of Electrical Engineering and Computer Science at the University of California at Berkeley, where he has been since 1999. Prior to that, he was with the University of Illinois at Urbana-Champaign from 1993 to 1999, and was at AT\&T Bell Laboratories from 1984 to 1990. His current research interests include distributed signal processing algorithms for wireless sensor and ad hoc networks, multimedia and peer-to-peer networking, multi-user information and communication theory, and wavelets and multi-resolution signal and image processing. Prof. Ramchandran is a Fellow of the IEEE. His research awards include the Elaihu Jury award for the best doctoral thesis in the systems area at Columbia University, the NSF CAREER award, the ONR and ARO Young Investigator Awards, two Best Paper awards from the IEEE Signal Processing Society, a Hank Magnuski Scholar award for excellence in junior faculty at the University of Illinois, and an Okawa Foundation Prize for excellence in research at Berkeley. He is a Fellow of the IEEE. He has published extensively in his field, holds 8 patents, serves as an active consultant to industry, and has held various editorial and Technical Program Committee positions.
\end{biography}

\begin{biography}{Yunnan Wu} (S'02-M'06) received the Ph.D. degree from Princeton University
in January 2006. Since August 2005, he has been a Researcher at Microsoft
Corporation (Redmond, WA, USA). His research interests include networking,
graph theory, information theory, game theory, wireless communications, and multimedia.
He was a recipient of the Best Student Paper Award at the 2000 SPIE
and IS\&T Visual Communication and Image Processing Conference, and
a recipient of the Student Paper Award at the 2005 IEEE International
Conference on Acoustics, Speech, and Signal Processing. He was awarded
a Microsoft Research Graduate Fellowship for 2003-2005.
\end{biography}

\begin{biography}{Changho Suh} received the B.S. and M.S. degrees in electrical engineering from Korea Advanced Institute of Science and Technology, Daejeon, Korea, in 2000 and 2002, respectively. Since 2006, he has been with the Department of Electrical Engineering and Computer Science in the University of California at Berkeley. Prior to that, he had been with the Department of Telecommunication R\&D Center, Samsung Electronics, Korea. His research interests include information theory and wireless communications.
He is a recipient of the Best Student Paper Award from IEEE International Symposium on Information Theory 2009 and the Outstanding Graduate Student Instructor Award in 2010. He awarded several fellowships: the Vodafone U.S. Foundation Fellowship in 2006 and 2007; Kwanjeong Educational Foundation Fellowship in 2009; and Korea Government Fellowship from 1996 to 2002.

\end{biography}

\end{document}